\documentclass{article}
\usepackage{multicol}
\usepackage{cec1999,multicol,times}

\usepackage{epsfig}

\begin{document}
\pagestyle{empty}
\sloppy

%\input{psfig.sty}
%\input epsf

% Some macros %
\def\s{\scriptstyle}
\def\ss{\scriptscriptstyle}
\def\cl{\centerline}
\def\ds{\displaystyle}
\newcommand{\be}{\begin{eqnarray}}
\newcommand{\ee}{\end{eqnarray}}
\newcommand{\nn}{\nonumber}

\def\lra{\longrightarrow}
\def\slashxi{{\xi}\!\!\!/} 
\def\nx{n(\xi,t)}
\def\nxi{n(\xi,t+1)}
\def\o{\over}
\def\C{{\s C}}
\def\xL{\xi_{\ss L}}
\def\xR{\xi_{\ss R}}
\def\pmic{P(\C_i,t)}
\def\pmicj{P(\C_j,t)}
\def\pmicl{P(\C_i,t+1)}
\def\pmico{P(\C_i,0)}
\def\pmicf{P'(\C_i,t)}
\def\pmicjf{P'(\C_j,t)}
\def\pmicc{P_c(\C_i,t)}
\def\pmicjc{P_c(\C_j,t)}
\def\pmiclf{P'(\C_l,t)}
\def\pmicLf{P'(\C_i^{\ss L},t)}
\def\pmicRf{P'(\C_i^{\ss R},t)}
\def\px{P(\xi,t)}
\def\pxl{P(\xi,t+1)}
\def\pxo{P(\xi,0)}
\def\pxf{P'(\xi,t)}
\def\pnxf{P'(\slashxi_i,t)}
\def\pxc{P_c(\xi,t)}
\def\pnxc{P_c(\slashxi_i,t)}
\def\pnx{P(\slashxi_i,t)}
\def\pxLf{P'(\xi_{\ss L},t)}
\def\pxRf{P'(\xi_{\ss R},t)}
\def\pxL{P(\xi_{\ss L},t)}
\def\pxR{P(\xi_{\ss R},t)}
\def\delpx{\delta P(\xi,t)}
\def\delpxl{\delta P(\xi,t+1)}
\def\delpxo{\delta P(\xi,0)}
\def\fav{{\bar f}(t)}
\def\fxi{{\bar f}(\xi,t)}
\def\frat{{\fxi\over \fav}}
\def\fx{{f_{\xi}}}
\def\fxL{{f_{\xi_L}}}
\def\fxR{{f_{\xi_R}}}
\def\fmic{f(\C_i,t)}
\def\fmicj{f(\C_j,t)}
\def\fmicl{f(\C_l,t)}
\def\fmicrat{{\fmic\o\fav}}
\def\fmicratj{{\fmicj\o\fav}}
\def\fmicratl{{\fmicl\o\fav}}
\def\dfx{\delta\fx}
\def\dfxL{\delta\fxL}
\def\dfxR{\delta\fxR}
\def\dfxi{\delta\fxi}
\def\pop{{\cal A}_p}
\def\str{{\cal A}_s}
\def\muti{{\cal P}({\ss C_i})}
\def\mutij{{\cal P}({\ss C_i\ra C_j})}
\def\mutji{{\cal P}({\ss C_j\ra C_i})}
\def\mutis{{\cal P}({\s \xi})}
\def\mutijs{{\cal P}({\s \slashxi_i\ra\xi})}
\def\crossij{{\cal C}_{\ss C_iC_j}^{(1)}(k)}
\def\crossjl{{\cal C}_{\ss C_jC_l}^{(2)}(k)}
\def\hamij{d^H(i,j)}
\def\hamijL{d^H_L(i,j)}
\def\hamijR{d^H_R(i,j)}
\def\hamilL{d^H_L(i,l)}
\def\hamilR{d^H_R(i,l)}
\def\ra{\rightarrow}
\def\p{{\tilde p}}
\def\ef{f_{\ss\rm eff}(\xi,t)}
\def\delef{\delta\ef}
\def\e{{\rm e}}
 
\twocolumn[

\title{``EFFECTIVE'' FITNESS LANDSCAPES FOR EVOLUTIONARY SYSTEMS} 

\vspace{0.1in}

\begin{center}{{\bf C. R. Stephens} }\\
NNCP, Instituto de Ciencias Nucleares, \\   
UNAM, Circuito Exterior, A.Postal 70-543 \\ 
M\'exico D.F. 04510 \\
e-mail: stephens@nuclecu.unam.mx \\
\end{center}

\vspace{0.25in}
]

%\twocolumn{

\begin{abstract} 

In evolution theory the concept of a fitness landscape has played
an important role, evolution itself being portrayed as a hill-climbing
process on a rugged landscape. In this article it is shown that in
general, in the presence of other genetic operators such as mutation
and recombination, hill-climbing is the exception rather than the rule.
This descrepency can be traced to the different ways that the 
concept of fitness appears --- as a measure of the number of fit 
offspring, or as a measure of the probability to reach reproductive
age. Effective fitness models the former not the latter and
gives an intuitive way to understand population dynamics as flows on an 
effective fitness landscape when genetic operators other than selection
play an important role. The efficacy of the concept is shown using
several simple analytic examples and also some more complicated
cases illustrated by simulations. 

\end{abstract}

\section{Introduction}

The notion of a fitness landscape, as originally conceived by 
Sewell Wright \cite{wright}, has played an important role as a unifying
concept in the theory of complex systems in the last ten years or so.
In particular, Stuart Kauffman \cite{kauffman} has utilized the 
concept for addressing the issue of the ``origin of order'' in the 
biological world, offering another paradigm for such order --- spontaneous
ordering --- as opposed to the traditional Darwinian view of ``order
by selection.'' 

In evolutionary computation the concept of a fitness landscape has not played 
the same sort of central role, although that is not to say it has not
played an important role (see for example \cite{fogel1} for an historical
perspective and \cite{jones} for a more recent account of the role of
landscapes in evolutionary computation). Given that the latter
treats very much the same type of system, at least mathematically, as
population genetics it should be clear that a closer scrutiny of 
population flows on fitness landscapes 
in evolutionary computation will help us understand how and why 
systems such as genetic algorithms (GAs) behave the way they do, and in
particular to understand what effective degrees of freedom are being
utilized in their evolution. One reason why it has not played such a role is that
evolutionary computation is somewhat ``simulation driven.'' Paradigmatically 
simple landscapes, such as the needle-in-the-haystack (NIAH) landscape 
considered by Eigen \cite{eigen} do not generate a great deal of
interest. However, landscape analysis is usually so difficult that one
has to typically start at the level of simple models. In terms of
population dynamics on fitness landscapes much attention has been paid to
adaptive walks on the hypercubic configuration spaces of the 
Kauffman $NK$-models \cite{kauffman}.
Such dynamics can be of interest biologically speaking, but do not seem to 
be of particular interest for evolutionary computation. Thus, there has been 
an ``expectation gap'' between what theoretical biologists, physicists, and
mathematicians have been able to achieve in landscape theory and what
the evolutionary computation community expects.

Landscape analysis in GA theory, for instance, has tended to focus on the 
relation between problem difficulty and landscape modality; the
assumption being that more modality signifies more difficulty. Obviously
a classification of landscapes into those that are difficult for an
evolutionary algorithm and those that are easy would be immensely useful.
As has been discussed however, \cite{goldfoga3}, \cite{fogel2}, things can be 
somewhat counter-intuitive. For instance, the NIAH landscape
is unimodal yet, as is well known, is difficult for a GA, and for that 
matter just about anything else. On the other hand
a maximally modal function, such as the porcupine function \cite{ackley}, can be
easy. The moral here is that modality is not the same thing as ruggedness. Of
more importance is the degree of correlation in the landscape. Correlation
structure is obviously of the highest importance in search as it is a rough
measure of the mutual information available between two points of the landscape.
The correlation length, $\xi$, of the landscape
is a characteristic measure of the degree and extent of such correlations. Points 
separated by a distance $>\xi$ will be essentially uncorrelated, while points
at a distance $<\xi$ will be substantially correlated. In the 
NIAH landscape the natural correlation length is zero, there
being no indication anywhere in the landscape of the existence of the isolated
global optimum. 

When talking about fitness landscapes it is important to 
distinguish between static and dynamic landscape properties.
The degree of ruggedness is a static concept. However, what is of importance
in biology, as well as evolutionary computation, is how a population flows
on a given landscape. In fact, one might argue that the whole problem of 
evolution can be understood from the point of view of flows on fitness landscapes.
Obviously, to specify a flow one has to specify a dynamics. There is then 
an important question: what properties are robust, i.e. universal, under 
a change in the dynamics and which are sensitive? It would clearly be of 
enormous interest to have a ``theory of landscapes,''
though this is an extremely ambitious task. The work of the ``Vienna'' 
group is of particular interest in this context (see for example
\cite{stadler} and references therein).
The majority of previous work has been restricted to dynamics generated 
by one or both of the two genetic
operators selection and mutation, some celebrated results being 
Fisher's fundamental theorem of natural selection \cite{fisher} 
and the concept of an error threshold \cite{eigen}. 
In particular, adaptive walk models for Kaufmann $NK$
landscapes \cite{kauffman} offer an arena wherein some analytical progress 
can be made. Work that includes the effect of recombination 
has been less forthcoming, especially in terms of analytical work.

A fundamental question is: how do the different genetic operators 
such as selection, mutation and recombination affect population flows? 
Here, I am defining a genetic operator to be any operation $H$ 
such that $P(t+1)=HP(t)$, where $P(t)$ is the population at time $t$. 
It is common to think of selection as not being a genetic operator
as it acts only on expressed behavior, i.e. the phenotype. However, given
that, at least formally, there exists a genotype-phenotype map 
selection also acts at the genotypic level.
A lot of the power behind the standard visualization of a fitness 
landscape has been associated with the view that evolution is a 
hill-climbing process on such a landscape. This intuition however 
is linked to a particular class of dynamics --- principally selection. 
Thus, one genetic operator dominates the intuition behind the landscape. 
In the presence of other genetic operators it is quite generic  
that population flows are not simple hill climbing processes. What one
requires is a more democratic approach that treats the various
genetic operators on a more equal footing. Thus, one is
led to enquire as to whether there exist other ways of thinking of
landscapes so as to restore an intuitive picture of landscape 
dynamics in the presence of genetic operators other than selection. 
A precedent for such an alternative type of landscape  
can be found in thermodynamics where one may consider the difference
between energy and free energy, the latter being the more relevant
quantity for determining the behavior of a system. In particular the
system's properties are much more readily seen from the 
free energy landscape than the energy landscape.   

The plan of this contribution is as follows: in section 2 I will discuss some
issues related to the definition of fitness drawing in particular
a strong distinction between the concepts of ``reproductive'' fitness and
``offspring'' fitness. In section 3, I discuss landscape statics
and dynamics showing that a realistic landscape is almost always 
explicitly time dependent. In sections 4 and 5, I present several, generic
analytic and numerical examples wherein the corresponding population
flows cannot be intuitively understood on the standard fitness landscape 
thus illuminating the need for an alternative paradigm. In section 6 this
paradigm is introduced and discussed and the examples of the previous
two sections reconsidered. Finally, in section 7, I draw some conclusions.  

\section{What is fitness?}

Mathematically it is quite simple to define fitness: $f_{\ss Q}:\lra R^+$, 
where $Q$ is the space of phenotypes and is of dimension $D_Q$. 
It is natural to define fitness as a
function on phenotypes given that it is the phenotype that manifests the 
physical characteristics on which natural selection acts. However, the 
raw material of an evolutionary system is the genotype. Hence, one needs to
know how fitness manifests itself at the genotypic level. For that 
one defines a genotype-phenotype map, $\phi:G\lra Q$, where $G$ is 
the space of genotypes which has dimension $D_G$.
One may thus define an induced fitness function on the space of genotypes, 
$f_{\ss G}=f_{\ss Q}\circ\phi$. As the genotype-phenotype map is more often 
than not non-injective (many-to-one) the function $f_{\ss G}$ will be degenerate,
many genotypes corresponding to the same fitness value. Thus, fitness
gives an equivalence relation on $G$, many genotypes being equivalent selectively.
A simple example of this would be the standard synonym ``symmetry'' 
of the genetic code. I will thus refer to the equivalence of a set of genotypes
under the action of selection (i.e. they're all equally fit) as a ``symmetry'' 
between them. Obviously, by definition, selection preserves this symmetry. 

Having defined fitness as a mathematical relation we must now understand it
conceptually, in particular in its relation to other genetic operators. In
its simplest form \cite{strickberger} fitness and selection are measured by the
number of fertile offspring produced by one genotype versus another. However, 
this is not the way it is normally portrayed in evolutionary computation, 
which follows the lead of population genetics, where it is a measure
of the probability that an individual survives to reproductive age \cite{hofbauer}.
Clearly the two concepts are quite different. The second is a property 
of an individual, in that it does not depend on other genotypes, even 
though the reproductive fitness function may reflect 
``environmental'' effects. In the first case we must ask: what type of
fertile offspring are left to the next generation? Genetically 
identical copies of the parents or what? Without thinking too much of the
biological realities proportional selection means selection for 
reproduction of certain parents, which in
the absence of genetic``mixing'' operators such as mutation and recombination
leads to the production of offspring genetically identical to their parents.
However, this idea of fitness does not take into account the important
effect the other genetic operators may have in determining the complete
reproductive success of an individual. In particular, the effect of the other
genetic operators, as shall be shown below using several model examples, can be such
that the population flows on the standard fitness landscape cannot be understood
with any degree of intuition. 
When the effects of such genetic operators are small, and selection is dominant,
it is quite likely that the two fitnesses are quite close numerically. However, to
take another extreme, neutral evolution where all genotypes have the same probability
to reach maturity, the two will be quite different. We will introduce later the
concept of an ``effective fitness'' that can encompass both limits of strong and
weak selection within the same function. I will denote the reproductive fitness
of a genotype by $f(\C_i,t)$ and its success in producing offspring by the 
offspring fitness, $f_{\ss\rm eff}$, which will be defined mathematically in section 6.

Having discussed fitness we now come to the idea of a fitness landscape.
The concept of fitness landscape is
already implicit in the above definition of fitness which assigns to every $g\in G$
a real, non-negative number $r\in R^+$. Thus, one can intuitively think of a
mountainous landscape where fitness is the height function above some  
$D_G$-dimensional hyperplane. Of course, the visualization of a typical landscape
such as an $N$-dimensional hypercube is somewhat less picturesque. Crucial to the
concept of a landscape is the idea of a distance function, without which
the landscape is shapeless as we cannot say which genotypes are closely
related and which are not. A very common distance function, particularly natural in the
case of mutation and selection, is the Hamming distance, $d_{ij}$, between two 
genotypes $\C_i$ and $\C_i$. If the fitness function is an explicit
function of time the corresponding landscape will be a dynamical not a static object.
Almost invariably, landscape analysis has been restricted to landscapes
associated with a static {\it reproductive} fitness. Such landscapes, as 
we shall see, are the exception rather than the rule. In particular, 
given that the number of offspring of a genotype will almost inevitably be
a function of time a landscape based on $f_{\ss\rm eff}$ will be an
explicit function of time.

\section{Landscape Statics and Dynamics}

As mentioned above, a fitness landscape is normally thought of 
as a static concept, the reproductive fitness
assigned to a given configuration being independent of time. It is clear that
in any real, biological system this is a crude simplification. Any realistic
biological landscape must be time dependent, at least over some time scale, 
due to the effects on fitness of the environment. This is most clear in the
concept of coevolution where changes in one species can affect the fitness 
of another. Under certain circumstances, however, and over certain
time scales, a static landscape may be a good approximation to the actual one.
In this case a key concept is the ruggedness of the landscape which one
can partially think of as being a measure of the density of local optima but,
more importantly, is a measure of the degree of correlation in the landscape. 
An associated concept, a complexity catastrophe, shows that there are limits
to the power of selection in the case of both very smooth and very rough 
landscapes. 

In evolutionary computation there are many situations, especially in 
global optimization such as the canonical traveling salesman problem, 
where the landscape is strictly static. However, even in this case a
time-dependent landscape emerges in a very natural way. 
Consider any microscopic configuration, $\C_i$, that
we can represent by a set $g$ of $N$ elements $\{g_k\}$, $k\in[1,N]$. Such a
configuration could represent, for example,
the genotype of an organism, or a possible solution to a combinatorial problem. 
The fitness of such a configuration, $f(\C_i)$, I assume to be independent of time.
Now, consider another configuration, $\C_i\neq \C_i$, of fitness $f(\C_i)$. 
We ask: what common characteristics do the two configurations have? If they have
$N_2$ elements in common, represented by a set $\{g_c\}\subset\{g_k\}$, then we may ask
whether we may assign a fitness to those common characteristics. This can be simply
done by defining
\be
f(\xi,t)= f(\C_i)P(\C_i,t)+f(\C_i)P(\C_i,t)
\ee
where $P(C,t)$ is the probability of finding the configuration $C$ at time $t$. 
Of course, the above is very familiar to people working in GAs as $\xi$ is just 
a schema. I mention it as it is also of fundamental
importance in biology. Why? Because except in a computer simulation one can never
keep track of the evolution of all microscopic configurations. Typically, what are
of interest are more coarse-grained variables. For example, the fitness of a species, $S$,
we can consider as
\be
f(S,t)=\sum_{\C_i\in S} f(\C_i)P(\C_i,t)
\ee
where the sum is over all those genotypes that correspond to the species.   
The moral here is that any coarse graining whatsoever will introduce a time 
dependence into the fitness function for the coarse grained effective degrees of
freedom. Thus, in general the concept of a dynamic landscape is of more relevance than
a static one.
In the case of both biology and evolutionary computation fitness as measured 
in terms of number of offspring will be time dependent, hence, any landscape
portrayal of this function will also be time dependent.  

We now come to the question of how to impose a dynamics on the fitness landscape.
A population ${\cal M}(t)\equiv \{g(t)\}\subset G$, where $\{g(t)\}$ is the set
of genotypes present in the population at time $t$, flows according to 
\be
{\cal M}(t+1)={\cal H}(\{f(\C_i)\},\{p_k\},t){\cal M}(t)
\ee
where ${\cal H}$ is an evolution operator that depends on the fitness landscape, 
$\{f(\C_i)\}$, and the set of parameters, $\{p_k\}$, that govern the other genetic
operators; e.g. mutation and recombination probabilities. There are very many
choices by which one can implement a population dynamics. 
A simple one, applicable in both biology and evolutionary computation, is
that of pure proportional selection which gives the following
equation for the mean number, $n(\C_i,t)$, of genotype $\C_i$
\be
n(\C_i,t+1)={(f(\C_i)/ \fav)}n(\C_i,t)
\ee
where I assume the reproductive fitness landscape to be time
independent. In this case it is clear how the population flows --- monotonically
towards the global optimum of the landscape (neglecting of course finite size
effects). It is precisely such intuitive flows, according to the gradient of the
landscape, that have lent such power to the concept of evolution as a flow on
a fitness landscape. It should be fairly clear, and will be shown explicitly
in section 4, that in this dynamics reproductive and offspring fitnesses are 
equal due to the fact that the only operator present is reproductive selection.

Another interesting limiting
case is that of the adaptive walk whereby one represents the entire population
by one genotype that jumps instantaneously to a one-mutant neighbor. This 
dynamics and the effect of landscape ruggedness has been well studied.
However, this limiting case of hill-climbing, although it has biological interest, is not 
so interesting from the point of view of evolutionary computation as adaptive
walks get stuck at local optima. A more general dynamics for proportional selection,
mutation, and one-point crossover can be described by the
equation \cite{stewael1,stewael2}
\be
\pmicl=\muti\pmicc
+ \sum_{\s \C_i\neq \C_i}\mutij\pmicjc\label{maseqtwo}
\ee
where the effective mutation coefficients $\muti$ and $\mutij$ 
represent the probabilities that the genotype $\C_i$ remains unmutated
and the probability that the genotype $\C_j$ mutates to the genotype $\C_i$
respectively. $P_c(\C_i,t)$ is the mean proportion of strings $\C_i$ at time $t$
after selection and recombination. Explicitly 
\be
\pmicc=\left(1-p_c{l-1\o N-1}\right)\pmicf\nn\\ 
+ {p_c\o N-1}\sum_{k=1}^{l-1}
P'(C_i^L(k),t)P'(C_i^R(l-k)\label{eqth}
\ee
where $\pmicf=(f(C_i,t)/\fav)\pmic$, $\fav$ being the average 
population fitness. $p_c$ is the crossover probability 
and $k$ the crossover point. The quantities $P'(\C_i^L(k),t)$ and $P'(\C_i^R(k),t)$ 
are defined analogously to $\pmicf$ but refer to the coarse grained variables, 
i.e. schemata, $\C_i^L$ and $\C_i^R$ which are 
the parts of $\C_i$ to the left and right of $k$ respectively. 
One can illustrate the content of the equation with a simple
example: $0110100|0101010010$. The crossover point is at $k=7$ hence 
$\C_i^L$, as a schema, has $N_2=l=7$ while $\C_i^R$ has $N_2=l=10$.
An analogous equation, identical in functional form, for the case of a general
schema $\xi$ can also be derived \cite{stewael1,stewael2}. 
In the rest of the paper we will consider the dynamics generated by this 
equation in its various limits.

\section{Effect of other genetic operators: analytic examples}

In this section I will consider more explicitly 
the effect of genetic operators other than reproductive
selection on the population flow on fitness landscapes in the context
of some simple, analytically tractable models. I have discussed
that the intuition behind the idea of fitness is to a large extent that 
of giving a ``reproductive edge'' to certain genotypes, i.e. that certain
genotypes give rise to more offspring than others. Also, that the intuition
behind population flows on a fitness landscape is that of hill-climbing. 
Let's think about this somewhat more critically in the light of some
interesting counter-examples.

I will first consider some simple one and two locus systems. Consider a
single gene with two alleles, $0$ and $1$, which have the same reproductive 
fitness value, $f$. In the absence of mutations both the reproductive fitness
and the offspring fitness are the same. 
In the infinite population case, or on the average in the finite population case, 
$\Delta P(t)=P(1,t)-P(0,t)$ is constant in time. Therefore
any initial deviations from homogeneity in the initial population will
be preserved. For non-zero mutation rate any initial inhomogeneity will
be eliminated by the effect of mutations. Thus, if $\Delta P>0$ one will
find that the offspring fitness of allele $0$ is greater than that of allele
$1$ until the deviation is eliminated. Thus, the effect of mutations will be to
bring the system into ``equilibrium,''
i.e. into the homogeneous population state. During this approach
to equilibrium the less numerous allele, $0$, is ``selected'' more
than the allele $1$ in that it leaves more offspring. If the mutation rates
for changing allele $1$ to allele $0$ and for changing allele $0$ to allele
$1$ are not equal but are $p_1$ and $p_2$ respectively then the
differences between reproductive fitness and offspring fitness are 
even more pronounced as can be seen by
\be
\Delta P(t+1)=(1-2p_2)\Delta P(t)+(p_1-p_2)P(0,t)
\ee
In this case ${\rm limit}_{t\ra\infty}\Delta P(t)\ra ((p_1-p_2)/(p_1+p_2))$

Now consider a two-locus system, once again with two alleles, $0$ and $1$.
The fitness landscape we will take to be: $f(00)=f(01)=1$, $f(11)=10$, 
$f(10)=0.1$. The fitness landscape in this case is only partially degenerate:
the states $00$ and $01$ having the same fitness value. However, although
the reproductive fitness values are the same the offspring fitness 
values, once again, are different. The degeneracy in this case is 
lifted by the effect of mutation as can be seen from the equations
\be
P(00,t+1)={f(00)\o \fav}(1-p)P(00,t)+p^2{f(11)\o\fav}P(11,t)\nn\\
+{p(1-p)\o \fav}(f(01)P(01,t)+f(10)P(10,t))
\ee
\be
P(01,t+1)={f(01)\o \fav}(1-p)P(01,t)+p^2{f(10)\o\fav}P(10,t) \nn\\
+{p(1-p)\o \fav}(f(00)P(00,t)+f(11)P(11,t)) 
\ee
For $p<1/2$, and starting with a random initial population, in terms of 
number of offspring the configuration $01$ will be preferred to $00$. It is not
hard to see why: the fittest configuration is $11$ and this can more easily
mutate to $01$ than to $00$. Thus, there is a population flow from $00$ to $01$
in spite of the fact that there is no gradient in the reproductive fitness
landscape to induce it. On the contrary, even if there were a gradient in the
direction from $01$ to $00$ if it were not too large the mutation induced flow
from $00$ to $01$ could overcome it, i.e. the population can flow downhill
against the reproductive fitness gradient!
Thus, there is a tendency for the system to evolve along a preferred direction
not because of selection constraints but because the system has preferred
directions of change in the face of {\it random} mutations. This is the
phenomenon of orthogenesis.

Naturally, this phenomenon encourages one to ask just when neutral evolution \cite{kimura}
is actually ``neutral.'' In the above case although the flow is 
reproductively neutral in the direction $00,\ 01$ the neutrality, or symmetry, 
is broken by the presence of non-neutral adjacent mutants.
For a flat fitness landscape where all genotypes have fitness $f$
\be
P(\C_i,t+1)=\sum_{j=1}^{2^N-1}P(\C_i,t)p^{d_{ij}}(1-p)^{N-d_{ij}}
\ee
For a homogeneous population the number of states Hamming distance $d_{ij}$
from $\C_i$ is ${}^NC_{d_{ij}}$ which implies that the offspring fitness 
is the same for all genotypes. Small deviations from 
homogeneity will be manifest in small differences between the reproductive
fitness and the offspring fitness which will gradually diminish 
as the population homogenizes. If 
the initial population, ${\cal M}(0)$, is homogeneous then it is preserved to be
so in the evolution --- on the average. The equations I am using here are 
of course mean field equations and as such do not capture finite size effects
that can play an important role in neutral evolution. If the
landscape only has a flat subspace then how well one can describe the 
population evolution as being neutral will depend on where the population
is located and, if located predominantly in the flat subspace, what is the
Hamming distance to states not within the subspace and what is the fitness
of those states. Pictorially, if one thinks of a bowl with a flat bottom
then the sides of the bowl with the largest gradient will attract the 
population most strongly. 
 
Another example is that of the NIAH landscape in the presence
of mutation and selection. The landscape is $f(\C_i)=f_0,\ i={\rm opt}$, 
$f(\C_i)=f_1,\ i\neq{\rm opt}$, ${\rm opt}$ being the optimum genotype. 
One can use as a measure of order in the population
the relative concentration of optimum genotype, $P(C_{\ss\rm opt},t)$;
and in particular in the long time limit, $P(C_{\ss\rm opt},\infty)$, 
where a steady state population is reached known as a quasi-species. 
In this case as is well known \cite{eigen} $P(C_{\ss\rm opt},\infty)$ monotonically 
decreases as a function of the mutation rate $p$ until a critical rate, $p_{\ss\rm cri}$,
is reached beyond which $P(C_{\ss\rm opt},\infty)=1/2^N$. It is important to
realize that the fitness landscape is constant throughout this behaviour, i.e.
for $p=0$ the entire population climbs the fitness peak until $\fav=f_{\ss\rm opt}$,
while for $p\geq p_{\ss\rm cri}$ the population is uniformly dispersed throughout
the entire landscape. Clearly no intuition about these very different types 
of population flow endpoints can be gleaned from the structure of the landscape
itself. In one limit selection dominates in the other limit it has no effect. 
Whether or not the population will ascend a fitness peak due to reproductive 
selection depends crucially on the presence of another genetic operator --- mutation.

Above, I considered only mutation to support the supposition that reproductive
fitness and offspring fitness can be very different in the presence of other
genetic operators. Similar considerations also apply to recombination. 
To take an extreme example of this consider the following simple two-locus 
system in the presence of selection and recombination, but not mutation, defined
by a fitness landscape: $f(01)=f(10)=0$, $f(11)=f(00)=1$. The steady state 
solution of the evolution equation (\ref{maseqtwo}) 
is $P(11)=P(00)={1\over2}\left(1-{p_c\over2}\right)$, $P(01)=P(10)={p_c\over4}$.
For $p_c=1$ one sees that half the steady state population is composed
of genotypes that have zero fitness. In this case we see that populations
can even flow against infinite fitness gradients! Note also that this fixed point of 
the dynamics is in fact a stable one. For the other two-locus system mentioned above 
$P(00,1)=(1-(9.9p_c/12.1))P(00,0)$, $P(01,1)=(1+(9.9p_c/12.1))P(01,0)$.
Hence, once again we see a degeneracy in reproductive fitness being broken 
by the effect of another genetic operator, in this case recombination,
with a consequent differential in offspring fitness.  

\section{Effect of other genetic operators: examples from simulations}

Having shown how genetic operators such as mutation and recombination
can drastically alter the directions populations flow on fitness 
landscapes in some simple analytic models I will now illustrate the 
same phenomenon in the case of some much more complex, analytically
intractable models. I will be brief in detail referring, where 
applicable, the reader to the original literature. 

I will consider first the evolution of two schemata 
of various defining lengths in strings of size $8$
in the case of a simple counting ones landscape in the presence of
selection and recombination, but without mutation \cite{foga5}. 
In the absence of crossover
there will certainly be no preference for one schema length versus another, i.e.
the offspring fitness as well as the reproductive fitness of any two-schemata 
is on average the same. To analyze this situation consider the quantity 
$M(l)$ where $M(l)\equiv (n_{\ss opt}(l)-n_{\ss opt}(8))/n_{\ss opt}(8)$. Here, 
$n_{\ss opt}(l)$ is the number of optimal $2$-schemata of defining length $l$ 
normalized by the total number of length $l$ 2-schemata per string, i.e. $9-l$. 
By optimal $2$-schemata we mean schemata containing the global optimum $11$.
$n_{\ss opt}(8)$ is the number of optimal $2$-schemata of defining length $8$.

A large population of $5000$ $8$-bit strings was considered. 
Figure 1 shows an average over $30$ different runs of $M(l)$ versus time with 
$p_c=1$. As mentioned, without crossover there is essentially no preference for 
schemata of a given length. Adding in crossover leads to a remarkable change: Schemata 
prevalence is ordered monotonically with respect to length but with the larger schemata
being favored. Thus, although there is no preference in terms of reproductive
fitness for one schema length versus another quite the contrary is true in terms
of offspring fitness.
\begin{figure}[h]
$$
\psfig{figure=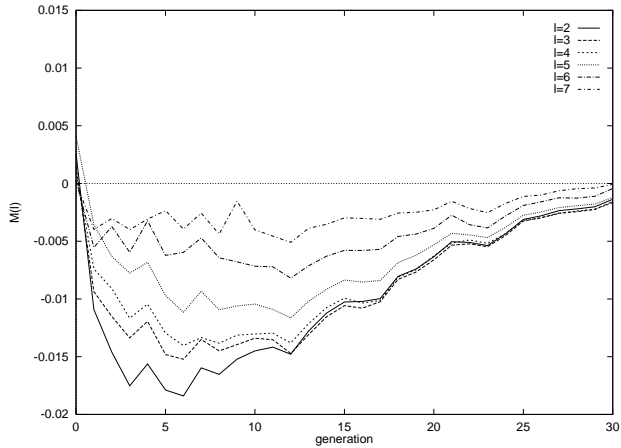,angle=-90,width=3.3in}
$$
\caption{Graph of $M(l)$ versus $t$ in unitation model with $p_c=1$.}
\end{figure}
Next, I will consider the differences between reproductive and offspring
fitness in the case of a simple auto-adaptive system. 
Specifically, one codes the mutation and $1$-point crossover probabilities 
into an $N_c$-bit binary extension of an $N$-bit genotype which is represented
by a non-degenerate fitness landscape, i.e. $f_G=f_Q$. This leads to a 
new $(N+N_c)$-bit genotype whose landscape has a degree of degeneracy of order 
$2^{N_c}$, 
i.e. the phenotype-genotype map is now $2^{N_c}$ fold degenerate. 

In practice, starting off with a random population,
where the average rates are $0.5$, one finds that the population in a class of 
interesting model landscapes self-organizes until preferred mutation and recombination
rates appear \cite{artlif}. Such self-organization cannot come about due
to any bias in terms of reproductive fitness, as by construction there is no such
bias. Neither can it come about by a spontaneous symmetry breaking (i.e. 
a spontaneous breaking of the genotype-phenotype degeneracy) due to stochastic
effects, i.e. finite size effects, as in the majority of the simulations the size
of the population was much bigger than $D_G$. What is happening is that even though
particular values for $p$ and $p_c$ are not selected for in terms of reproductive 
fitness they are selected for in terms of offspring fitness. Thus, in $G$ there is
a flow to a certain subregion of the space wherein the mutation and recombination
probabilities take on preferred values. Additionally, the probability distribution
associated with the various values of $p$ and $p_c$ narrows as time increases 
indicating that there is convergence of the population. 

As a specific example, consider a time-dependent landscape defined on 6-bit chromosomes 
that code the integers between $0$ and $63$. The landscape is time dependent in the
following way: the initial landscape has a global optimum situated at $10$ and 
$11$ and a local optimum at $40$ and $41$. However, after 60\% of the 
population reaches the global optimum the landscape changes instantaneously, converting
the original global optimum into a local one.  The original local
optimum at $40$ and $41$ remains the same but with a higher fitness value than the 
new local optimum at $10$ and $11$. Additionally, a new
global optimum appears at $63$. I will denote this landscape the ``jumper'' landscape. 
In Figure 2 one sees the results of an experiment where the mutation and 
crossover probabilities were coded either with three or eight bits to codify each
probability. Tournament selection of size 5 was used and a lower bound of $0.005$ 
for mutation imposed. The success of the self-adapting system in converging to the
time dependent global optimum was compared to that of an ``optimal'' fixed 
parameter system with $p=0.01$ and $p_c=0.8$.
\begin{figure}[h]
$$
\psfig{figure=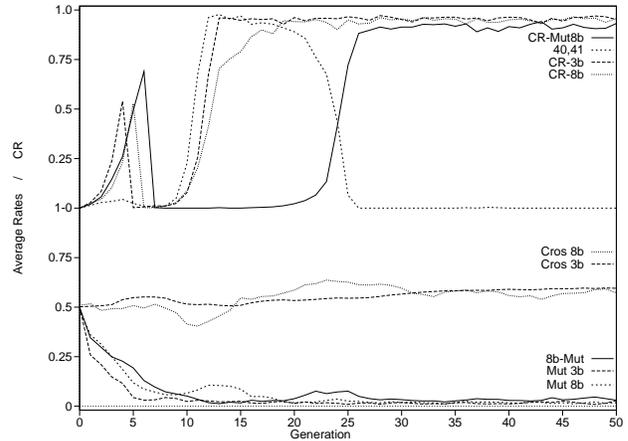,angle=-90,width=3.3in}
$$
\caption{Graph of relative concentration of the global optimum (CR) (upper graph)
and average crossover and mutation probabilities (lower graph) 
as a function of time for the ``jumper'' landscape. 
CR-3b and CR-8b are the results for $3$-bit and $8$-bit
encoded algorithms. CR-Mut8b is the result for coded mutation with $p_c=0$, 
with $40,41$ being the relative concentration of strings associated with the local
optimum at $40$ and $41$. Mut 3b, Mut 8b, Cross 3b and Cross 8b are the average
mutation and crossover probabilities in $3$-bit and $8$-bit representations. The
solid line for Mut8b is the average mutation rate in the case $p_c=0$.}
\end{figure}
The upper curves show the relative frequencies of the optima using $8$-bit and 
$3$-bit codification and also 
what happens when $p_c=0$ and only the mutation rate is coded. It is notable 
that the optimal fixed parameter system was 
incapable of finding the new optimum whereas the coded system had no such problem.  
For the case $p_c=0$ the curve $40,41$ shows the relative frequency
of the strings associated with the optimum at $40$ and $41$. Before the landscape
``jump'' this optimum is local, being less fit than the global optimum 
at $10$ and $11$. After the ``jump'' it is fitter but less fit than the new 
global optimum $63$ which is an isolated point. 

Notice that
the global optimum was found in a two-step process after the landscape change. First, the
strings started finding the optima $40$, $41$ before moving onto the true global
optimum, $63$. Immediately after the jump the effective population of the new
global optimum is essentially zero. The number of strings associated with 
$40$ and $41$ first starts to grow substantially at the expense of $10$ and
$11$ strings. At its maximum the number of optimum strings is still very low.
However, very soon thereafter the algorithm manages to find the optimum string which
then increases very rapidly at the expense of the rest. 

The striking result
here can be seen by comparing the changes in the relative frequencies with
the changes in the average mutation rate, especially in the case $p_c=0$. 
Clearly they are highly correlated.
First, while the population is ordering itself around the original optimum, there is
negative selection in terms of offspring fitness against high mutation rates as one can see
by the steady decay of the average mutation rate. After the jump there
is a noticeable increase in the mutation probability as the system now has to 
try to find fitter strings. As the global optimum is an isolated state it is much 
easier to find fit strings associated with $41$ and $40$. The population is
now concentrated on this local optimum and the average mutation rate 
decreases again only to find that this 
is not the global optimum, whereupon the average rate increases to aid the removal of the
population to the true global optimum. It is clear that there is a small delay
between the population changes and changes in the mutation rate. This 
is only to be expected given that there is no direct reproductive selective advantage 
in a given generation for a particular mutation rate. The selective advantage in
terms of the offspring fitness of a more mutable genotype over a less mutable one
arises via a feedback mechanism. 

The average mutation rate also grows due to another effect which is that 
the new optimum is more likely to be reached by strings with high mutation 
rates which then grow strongly due to their reproductive selective advantage. 
Thus high $p$ strings will naturally dominate the early evolution of the global
optimum. After finding the optimum however it becomes disadvantageous
in terms of offspring fitness to have a high mutation rate. Hence, low 
mutation strings will begin to dominate. Once again I emphasize, although 
there is no direct selective benefit in terms of reproductive fitness associated
with different mutation and crossover probabilities their ability to produce 
offspring that can adapt to the changing landscape is very different and this is
measured by the offspring fitness.

The final example concerns the size of giraffe necks \cite{virus}!
This model consists of a population of one thousand genotypes 
subject to random mutations. A genotype is a cellular automata with
binary elements which gives rise to a giraffe neck size, i.e. a phenotype, given by the 
number of automata elements that are ``switched on" at the  
fixed point (steady state) of the automata dynamics. As there are 
many different automata that can evolve to the same fixed point the 
genotype-phenotype mapping is highly degenerate. One ``master'' 
gene in particular plays a special role as it governs the way in which the
Boolean rules used in the evolution mutate. 

Each member of the population is selected for the next generation with
probability $P_i = f_i / \sum_j f_j$, where $f_i$ is the fitness
of phenotype $i$. Initially there are ample resources available from both 
small and large trees, the only selective criterion being that
giraffes prefer to choose a mate from among those that have similar
neck size. This ``social pressure" landscape is modeled by defining the
fitness of the $ i{\rm th} $ giraffe to be a function of its neck size 
$ n_i $ and the average neck size of the population, $ \left< n \right> $,
with value one if $ \left< n \right> -
\delta < n_i < \left< n \right> + \delta $ and zero otherwise.
Here, $ \delta > 0 $ is a tolerance window. Note that landscape
fitness depends only on neck size, hence all genotypes that correspond
to the same dynamical fixed point (phenotype) have the same reproductive
fitness. Thus, as in the other examples, there is no direct 
selective advantage for one genotype
versus another. To introduce time dependence into the landscape one
imposes a short period of drought in which food begins to be
available only in taller and taller trees. This period is mimicked by
making $ f_i = 1 $ if $\left< n \right> - \delta + \epsilon <
n_i < \left< n \right> + \delta + \epsilon $, where $ \epsilon $ is a
stress parameter, and zero otherwise. After this drought the landscape
is restored to its original state.

The ``master'' gene divides the population into two genetic categories, type
zero and type one, which can mutate one into the
other due to the effect of purely random mutations that have a probability 
$\mu$ except for the master gene which mutates at a rate $\nu$. 
Type zero chromosomes, by nature of the dynamical evolution
rules they are associated with, tend to
give rise to giraffe offspring with shorter necks, while type one chromosomes,
when they are expressed, tend to lead to giraffes with longer necks. Before
the draught there is a period in which type one is not expressed. At
a certain moment in time it becomes expressed then afterwards the drought starts.
The social pressure landscape implies there are two possible attractors: all
type one or all type zero. The effect of the drought is to change between one
and the other.

A typical experiment leads to the following results, the general behavior can
be seen in Figure 3:
\begin{figure}[h]
$$
\psfig{figure=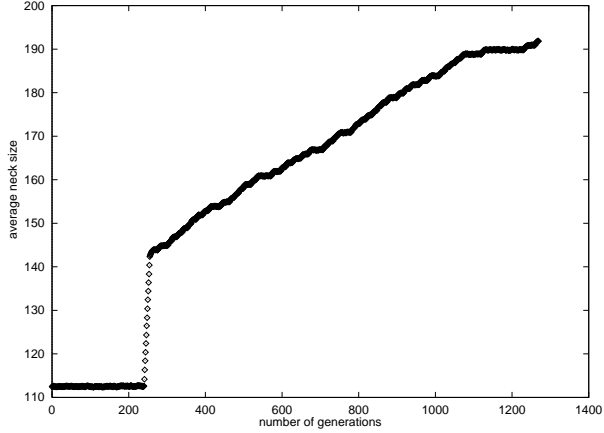,angle=-90,width=3.3in}
$$
\caption{ Graph of average giraffe neck size (in arbitrary units) as
a function of time for a population of 1000 giraffes. The drought starts at
generation 240 and lasts for 30 generations. Subsequent neck growth lasts
for another 1000 generations. The 
parameter values used were: $ \mu = 0.0025 $, $ \nu = 10^{-6} $, $ \delta =
2.0 $ and $ \epsilon = 1.0 $.}
\end{figure}
In the initial period of evolution, before the drought, average neck size is short.
After the drought arrives the average neck size grows very quickly.
After it ends it continues to grow, albeit more slowly, for
a substantial amount of time until a steady state is reached.  
These results can be explained quite simply:
in the period before the draught, and before expression, type one chromosomes
increase due to the effect of neutral drift. After expression they are
effectively selected against due to their tendency to produce giraffes with
longer necks that pass outside the tolerance threshold and therefore cannot 
reproduce. Thus, before the drought the offspring fitness of type one chromosomes
is low. However, due to the effect of mutations type one chromosomes 
are not eliminated totally but constitute about $1- 5\%$ of the total population.
After the draught starts the offspring fitness of type one chromosomes
increases substantially, given that they lead to giraffes of longer necks. The
result is that the population becomes dominated by type one chromosomes, 
with a small fraction of type zero remaining due to the effects of mutation.
After the end of the drought, as type one chromosomes tend to produce longer
necks, the average neck size increases until a steady state is reached and
it cannot grow anymore. 

In this model there is absolutely no direct 
reproductive selective difference between type one and type zero
chromosomes. The only advantage of one versus the other is in how they
produce well adapted offspring, as measured by the offspring fitness. 

\section{Effective fitness and effective fitness landscapes}

The previous two sections showed that it is difficult
to intuitively understand population flows on fitness
landscapes when genetic operators other than selection play an important role.
This is manifest in the fact that the most
selectively fit individuals reproductively do not always give 
rise to the most offspring. Thus, under many circumstances the 
hill-climbing analogy is not a good one. I emphasize that I am not 
just talking about a pathology that in practice is very unlikely to occur,
but rather a phenomenon that is central to both biology and evolutionary
computation whenever genetic operators other than pure selection are of
primary importance.

I have throughout constantly emphasized the difference between 
reproductive ``landscape'' fitness and offspring fitness. I now wish 
to define mathematically the latter. There are several possibilities
for such a quantity. For instance, based on the thermodynamic analogy 
\cite{leut} between an inhomogeneous two-dimensional spin model and 
a population evolving with respect to selection and mutation, one could define 
quite naturally the free energy per row as an effective measure of fitness.
Here, however, I will use another definition, more directly related to the 
traditional idea of fitness in biology and evolutionary computation. I will
define the offspring fitness as the ``effective fitness'' given by 
\cite{stewael1,stewael2}
\be
P(\xi,t+1)={f_{\ss\rm eff}(\xi,t)\o \fav}P(\xi,t)\label{efffit}
\ee
One may think of the effective fitness as representing the effect of all
genetic operators in a single ``selection'' factor. 
$f_{\ss\rm eff}(\C_i,t)$ is the fitness value at time $t$ required 
to increase or decrease $P(\C_i,t)$ by pure selection by the same amount 
as all the genetic operators combined in the context of a reproductive fitness $f(\C_i)$. 
If $f_{\ss\rm eff}(\C_i,t)>f(\C_i,t)$ the effect of the genetic
operators other than selection is to enhance the number present of genotype
$\C_i$ relative to the number found in their absence.
Obviously, the converse is true when $f_{\ss\rm eff}(\C_i,t)<f(\C_i,t)$.
The exact functional form of the effective fitness obviously depends on
the set of genetic operators involved. For the fairly general case of
equation (\ref{maseqtwo}) we have
\be
f_{\ss\rm eff}(\C_i,t)= \nn \\ 
\left({\muti\pmicc + 
\sum_{\s \C_i\neq \C_i}\mutij\pmicjc\o P(\C_i,t)}\right)\fav\label{effittwo}
\ee
Note that it depends on the actual composition of the 
population. 

Let us see how the concept of effective fitness and an effective fitness
landscape can help us better understand the results of section 4. 
In the simple one-locus model there; in the absence of mutations
$f_{\ss\rm eff}(\C_i,t)=f(\C_i,t)=f,\ \ \forall i=0,1$. For $p\neq 0$
\be
f_{\ss\rm eff}(\C_i,t)=\left((1-2p)+(p/P(\C_i,t))\right)f
\ee
Thus, we can see explicitly that when the population is homogeneous 
effective fitness is the same as reproductive fitness. Deviations
from homogeneity result in a higher effective fitness for the less
numerous genotype. The effective fitness advantage then decreases as the
system approaches equilibrium. The effective fitness landscape for
this model as a function of time can be seen in Figure 4.
\begin{figure}[h]
$$
\psfig{figure=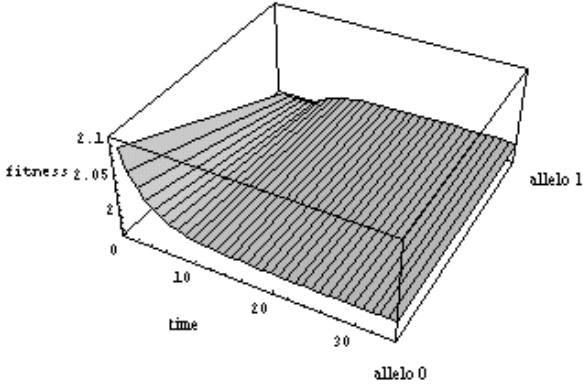,angle=-0,width=4.5in}
$$
\vspace{-2.7truein}
\caption{Graph of effective fitness as a function of time for the one-locus
model of section 4.}
\end{figure}
In this case, $f=2$, $p_1=0.15$, $p_2=0.1$ and $P(0,0)=P(1,0)=0.5$.
Note the significant differences in $f_{\ss \rm eff}(i,0)$ as a result
of which there is a population flow from $1$ to $0$ along this effective
fitness gradient even though there is no reproductive fitness gradient. 
This gradient decreases monotonically as a function 
of time thus the effective landscape becomes flat asymptotically. We can see from
(\ref{efffit}) that this will be a generic property of any effective fitness 
landscape, the only steady state solutions of the equation being
$f_{\ss\rm eff}(\C_i,t)=\fav\ \forall \C_i$, or $P(\C_i,t)=0$. Thus, already
we see the usefulness of effective fitness --- it indicates by its
deviation from flatness how close the population is to a steady state.

For the two-locus model the reproductive fitnesses of $00$ and $01$ are
equal. However, their effective fitnesses are quite different being:
$f_{\ss\rm eff}(00,0)=(1-0.9p+9.9p^2)$ and $f_{\ss\rm eff}(01,0)=(1+9p-9.9p^2)$  
respectively, where initial proportions of all four states are taken to be equal. 
The effective fitness landscape in this problem is shown in Figure 5 for $t=0$ with 
$p=0.05$, $P(00,0)=3$, $P(01,0)=0.2$, $P(10,0)=0.4$ and $P(11,0)=0.1$.
\begin{figure}[h]
$$
\psfig{figure=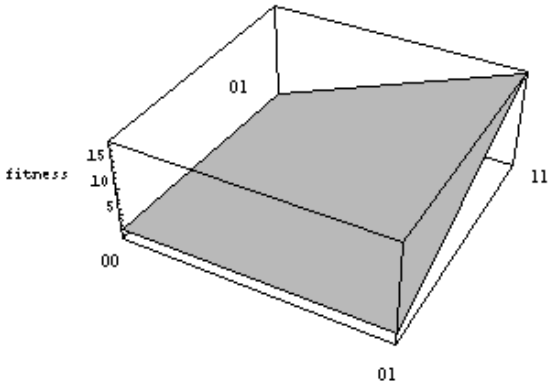,angle=-0,width=4.5in}
$$
\vspace{-2.7truein}
\caption{ Graph of effective fitness at $t=0$ for the two-locus
model of section 4.}
\end{figure}
Notice the initial effective fitness gradient from $00$ to $01$ 
that $\ra 0$ as $t\ra\infty$ as can be seen in Figure 6 which shows
the same effective fitness landscape as in Figure 5 but at late times.
Once again note the flatness of the effective landscape at late times.
\begin{figure}[h]
$$
\psfig{figure=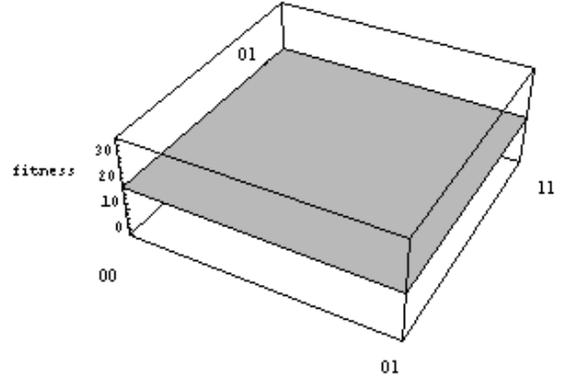,angle=-0,width=4.5in}
$$
\vspace{-2.7truein}
\caption{ Graph of effective fitness at late times for the two-locus
model of section 4.}
\end{figure}
 
In the case of a strictly flat fitness landscape characteristic
of neutral evolution the effective fitness is
\be
f_{\ss\rm eff}(\C_i,t)=f\sum_{j=1}^{2^N-1}{P(\C_j,t)\o P(\C_i,t)}
p^{d_{ij}}(1-p)^{N-d_{ij}}
\ee
For a homogeneous population $f_{\ss\rm eff}(\C_i,t)=f \forall \C_i,t$.
Thus, under these circumstances the effective fitness landscape is as flat
as the normal one. Small deviations from 
homogeneity will be manifest in small corrugations of the effective fitness
landscape which will gradually diminish as the population homogenizes. If the
landscape only has a flat subspace the system will seek to escape the flat
subspace by way of the direction with the highest effective fitness gradient. 

In the case of the NIAH landscape the effective fitness of the optimum string is
\be
f_{\ss\rm eff}(C_{\ss\rm opt},t)=f_0(1-p)^N \nn \\ 
+f_1\sum_{\C_i\neq \C_i}
{P(\C_i,t)\o P(C_{\ss\rm opt},t)}p^{d_{ij}}(1-p)^{N-d_{ij}}
\ee
In Figure 7 we see a plot of the effective fitness as found in a simulation of the
Eigen model in the steady state as a function of $p$. In this case $f_0=9$ and
$f_1=1$. 
\begin{figure}[h]
$$
\psfig{figure=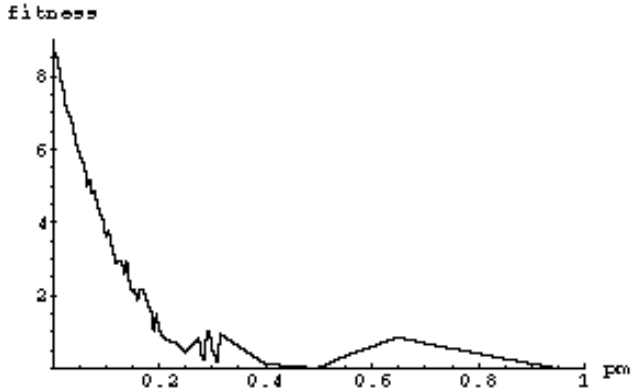,angle=-0,width=3.3in}
$$
\vspace{-0.52truein}
\caption{ Graph of effective fitness versus $p$ in the steady state limit of the
NIAH landscape.}
\end{figure}
For $p=0$, 
$f_{\ss\rm eff}(C_{\ss\rm opt},\infty)=f_0$ i.e. reproductive and effective
fitness are the same. Note how the error threshold, located at about $0.2$,
manifests itself in terms of the effective fitness --- that at and above the 
threshold $f_{\ss\rm eff}(C_{\ss\rm opt},t)\ra \fav\approx f_1$. i.e. once again
the effective fitness landscape will be flat. The curve for $p>p_{\ss\rm cri}$ 
is quite noisy due to the fact that there are very few optimum strings in the 
population. Thus, the 
effective fitness itself can serve as an order parameter to distinguish the
selection dominated regime from the mutation dominated one. 

Consider now the examples of section 5. In the first example, in Figure 8 we see
the effective fitness as a function of time for the various two-schemata. 
Note the initial monotonic ordering of the schemata --- the longest being the 
most effectively fit. 
\begin{figure}[h]
$$
\psfig{figure=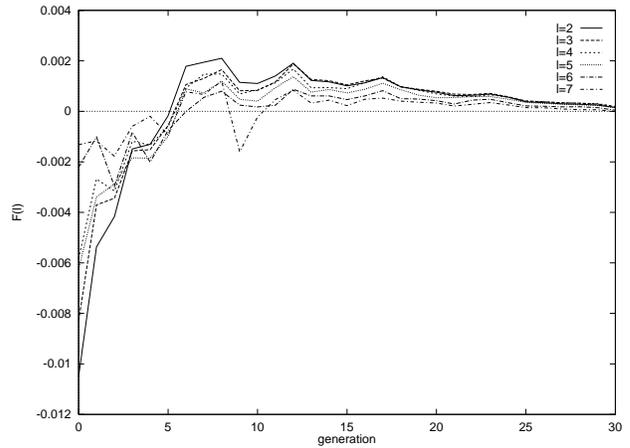,angle=-90,width=3.3in}
$$
\caption{ Graph of effective fitness versus time for the unitation model of Figure 1.}
\end{figure}

For the self-adaptive system if we concentrate for the moment on just selection and
mutation the effective fitness of a given string is
\be 
f_{\ss\rm eff}(\C_i,t)=f(\C_i,t)(1-p_i)^N \nn \\
+\sum_{\C_i\neq \C_i}p_j^{d_{ij}}(1-p_j)^{N-d_{ij}}f(\C_i,t){P(\C_j,t)\o P(\C_i,t)}
\label{adapt}
\ee
where $p_i$ is the mutation rate associated with the genotype $\C_i$. 
To understand the behavior shown in Figure 2 we first realize that there
are two contributions to $f_{\ss\rm eff}(\C_i,t)$: one from the genotype $\C_i$
itself and another from all the other genotypes that can mutate to it.
For the first contribution we can see that genotypes with low mutation rates
will be preferred while for the second contribution genotypes with
higher mutation rates will tend to be preferred. Thus, we can see a quantitative
way of evaluating the explanation given in section 5. Initially, before the
landscape jump, selection dominated in that the first term in (\ref{adapt}) was
the most important. Consequently the average mutation rate decreased as genotypes 
of low $p_i$ were more effectively fit. After the jump there is at first 
a preference for the local optimum at $40$, $41$. This is due to the fact that
both terms in the effective fitness are playing an important role whereas
for the needle-like global optimum, as the original population before the 
jump had converged to $10$, $11$, there was no contribution from the first
term to $f_{\ss\rm eff}(63,t)$. The fact that the average Hamming 
distance from $10$ and $11$ to $63$ is $3.5$, while from $10$, $11$ to $40$, 
$41$ it is $2.75$, explains why the system first converges to $40$, $41$ 
before passing on to the global optimum. It is clear too why the average
mutation rate increases at two distinct points in the the search process. 
It is precisely when the second contribution in the effective fitness 
is playing a dominant role in aiding the search.

The giraffe example is rather similar to that of the self-adapting system, 
the master gene being analogous to the part of the genotype that
codes for the mutation and recombination rates in the latter. It plays
no role in reproductive selection but does play an important role in
effective selection type one chromosomes having a higher effective 
fitness than type two during and after the drought. 

All these examples show clearly how different the effective fitness landscape
is from the reproductive fitness landscape and how population flows can 
be understood much better within the framework of the former.
 
\section{Conclusions}

In evolutionary computation, as in biology, it is obviously of crucial importance
to know what generic properties the effective degrees of 
freedom exploited by a genetic system possess. Of particular concern 
is what constitutes a ``fit'' string. I have shown that in many
situations high/low landscape reproductive fitness does not correspond to high/low 
effective fitness and it is the latter that really determines reproductive
success. 

Several criticisms of the effective fitness concept come 
immediately to mind. First of all, why
do we need another type of landscape? Isn't the present one good enough? 
After all, the notion of an adaptive landscape has turned out to be one of the most 
powerful concepts in evolutionary theory. One of the strongest motivations
for the fitness landscape concept is that it allows one to intuitively 
understand population flows as hill climbing processes, thus offering a compelling
paradigm for how evolution works. However, as I have demonstrated the hill 
climbing analogy is intimately linked to a certain type of dynamics associated
with pure selection. In the presence of other genetic operators hill climbing
is the exception rather than the norm and so the idea of a fitness landscape
thereby loses some of its appeal. Another possible criticism is that 
effective fitness is intrinsically time-dependent. Doesn't this lead 
to an extra level of complication relative to that of a static landscape? 
As I have argued there are several motivations for thinking of a time dependent 
landscape as being more fundamental than a static one. First of all, it is
more biologically realistic as environmental effects that affect 
fitness are almost inevitably time dependent. Secondly, even if a landscape
is static in terms of the microscopic degrees of freedom it will be time
dependent in terms of any coarse grained degrees of freedom.
On might also argue that the definition is tautological --- survival of the 
survivors. Such circularity is not very different to that which appears in
other sciences such as physics where one can level the same sort of
criticism at an equation such as Newton's Second Law. Such circularity is 
usually a hallmark of a truly fundamental concept and
its definition. Additionally, in the present context we have a way to, 
at least in principle, explicitly compute it in terms of the parameters
associated with the various genetic operators.  

Another advantage of the effective fitness concept is that it allows one
to quantitatively understand the different mechanisms
by which order may arise in evolution. i.e. that ``order'' may arise due
to the effect of genetic operators other than selection. In particular it provides a 
framework within which neutral evolution and natural selection can be
understood as different sides of the same coin, and in particular under
what circumstances neutral mutations may lead to adaptive changes. In this
sense the phenomenon of orthogenesis, i.e. genetic drives in the presence 
of random mutations, is nothing more than the appearance
of an effective fitness gradient in the case where there was no original reproductive 
fitness gradient. Clearly, much more work needs to be done in understanding
the effective fitness concept and developing an intuition for landscape
analysis based on it rather than the traditional idea of fitness landscape.
 
\subsubsection*{Acknowledgements} 

This work was partially supported through DGAPA-UNAM grant number IN105197.
Many of the basic ideas presented here were developed in collaboration 
with Henri Waelbroeck to whom I am grateful for many stimulating
conversations. I also wish to thank Sara Vera for help with the graphs
and David Fogel for useful comments on the manuscript. 

\vspace{-0.2truein}

\subsubsection*{}

\end{document}